\documentclass{article}  
\usepackage{bigsky2009}
\usepackage{graphicx}

\newcommand{\mean}[1]{\left\langle #1 \right\rangle}

\frompage{000} \topage{000}                                              
\def\rightmark{Is the Quark Gluon plasma strongly coupled?}
\def\leftmark{Lacey, Taranenko and Wei}
 
\title{Is the quark gluon plasma produced in RHIC collisions strongly coupled?} 
\authors{
{Roy A. Lacey$^1$, Arkadij Taranenko$^1$ and Rui Wei $^{1}$ 
}\\[2.812mm]
{\normalsize
\hspace*{-8pt}$^1$ Chemistry Dept., Stony Brook University \\ 
Stony Brook NY 11794-3400, USA\\[0.2ex] 
}}
 
\abstract{
          Recent hexadecapole (${v_{4}}$) and elliptic (${v_{2}}$) 
flow measurements are used to constrain estimates 
for the degree of local equilibrium, mean free path $\lambda$, and the viscosity to 
entropy density ratio ($\frac{\eta}{s}$) of the plasma produced in Au+Au collisions 
at $\sqrt{s_{NN}}~=~200$ GeV. The eccentricity-scaled flow coefficients $\frac{v_{2}}{\varepsilon_{2}}$ and 
$\frac{v_{4}}{\varepsilon_{4} }$ indicate that the plasma achieves a degree of 
local equilibrium within $\sim 5 - 10$\% of the value expected for a fluid with $\frac{\eta}{s}$ 
equal to the conjectured lower bound of $1/4\pi$. Estimates for $\lambda$ and $\frac{\eta}{s}$ 
as a function of collision centrality and particle transverse momentum $p_T$, points to 
transverse expansion dynamics compatible with a strongly coupled low viscosity plasma. 
}

 
\begin{document}
 
\maketitle
\setcounter{page}{1}

\section{Introduction}\label{intro}

In central and mid-central Au+Au collisions at the Relativistic Heavy ion Collider (RHIC),
matter is produced at energy densities well in excess of the value ($\sim 1$~GeV/fm$^3$) 
required for a de-confinement transition to the quark gluon plasma (QGP). 
A strong indication that such a de-confinement transition indeed occurs, is the important role that 
quark-like degrees of freedom have been found to play in the transverse expansion dynamics 
leading to anisotropic flow. Such flow is routinely characterized  
by the even order Fourier coefficients  
\[
v_{\rm n} = \mean{e^{in(\phi_p - \Phi_{RP})}}, {\,\,} n=2,4,...\;,
\]
where, $\phi_{p}$ is the azimuthal angle of an emitted particle, 
$\Phi_{RP}$ is the azimuth of the reaction plane and the brackets denote 
averaging over particles and events. The second ($v_2$) and fourth ($v_4$) order 
coefficients characterize the magnitude of elliptic and hexadecapole flow 
respectively.

	At the highest RHIC collision energy of $\sqrt{s_{NN}}=200$\,GeV,  
a universal scaling of elliptic flow, suggestive of constituent quark-like degrees 
of freedom in the collision zone, has been discovered  \cite{Adare:2006ti,Lacey:2006pn} 
and is now well established for a broad range of collision centralities, particle species 
and transverse kinetic energies.
For transverse momenta $p_T{\,\mathop  < \limits_\sim\,}1.5$\,GeV/$c$, hydrodynamic calculations 
that model a locally equilibrated QGP (with little or no viscosity) show good agreement with 
the data \cite{Heinz:2001xi,Shuryak:2008eq,Romatschke:2007mq}. 
A recent transport calculation which incorporates gluon dynamics \cite{Xu:2007jv}, 
also indicate good agreement with the magnitude and trend of centrality dependent $v_2$ data.
Thus, the measurement and study of flow have been central to the confluence of 
experimental results 
which now bear evidence for the creation of the QGP in  
heavy ion collisions at RHIC. 

	A less settled question which is still debated intensely, is the degree to which the 
QGP is thermalized \cite{Borghini:2005kd}, and  whether it is strongly or weakly 
coupled \cite{Shuryak:2008eq,Asakawa:2006tc}. Experimental constraints which 
allow estimates of the ratio of viscosity to entropy density $\frac{\eta}{s}$
and the mean free path $\lambda$, are crucial ingredients to the resolution of this question.
In this contribution we show that the combined use of double differential $v_2$ and $v_4$ measurements 
provide such constraints, and consequently, lend new and important insight to 
this question.
%

\section{Flow, local equilibrium \& the coupling strength of the QGP}\label{techno}  

	The use of flow correlations as a probe for the nuclear equation of state (EOS) and 
the transport coefficients of hot and dense nuclear matter has been recognized for 
quite some time \cite{GlassGold_AP6,Chapline_PRD8,Sheid_PRL32,Stocker_PRL44}. 
The connection is made transparent in the framework of perfect fluid
hydrodynamics where the conceptual link between the conservation laws (baryon number,
and energy and momentum currents) and the fundamental properties of a fluid (its equation of
state and transport coefficients) is straightforward. 
Much current effort to understand and determine transport properties are focused on 
several microscopic models (for recent reviews see for example 
Refs.~\cite{Shuryak:2008eq,Schaefer:2009dj}).
Hybrid approaches which involve the parametrization of deviations from hydrodynamic 
behavior (i.e full local equilibrium) are currently being studied 
as well~\cite{Bhalerao:2005mm,Drescher:2007cd,Gombeaud:2007ub}. The latter exploits 
the fact that results from the  Boltzmann equation reduce to those from perfect fluid 
hydrodynamics when the mean free path $\lambda$ becomes small~\cite{Marle69}.

\subsection{Deviation from local equilibrium}\label{equilibrium}

	Further insight on the extent to which a system deviates from full local equilibrium 
can be obtained via simultaneous study of the scaling violations of $v_2$ and $v_4$. 
Here, the central idea is that partial equilibrium  breaks the scale invariance of 
perfect fluid hydrodynamics (which requires full local equilibrium), 
and thus, gives rise to specific measurable scaling 
violations \cite{Bhalerao:2005mm,Drescher:2007cd,Song:2008si}.
The quantification of such scaling violations can then be used to constrain 
an estimate of the ``degree'' of local equilibrium and the transport coefficients.

	One such violation of the eccentricity-scaled second harmonic $v_2/\varepsilon_2$,  
has been recently parametrized \cite{Bhalerao:2005mm,Drescher:2007cd} via the Knudsen 
number $K ={\lambda}/{\bar R}$ \cite{Knudsen:1909}, as; 
\begin{equation}
 \frac{v_{2}}{\varepsilon_2} = \frac{v_{2}^{\rm{h}}}{\varepsilon_2} 
 \frac{K^{-1}}{K^{-1}+K_0^{-1}},
\label{eq2}
\end{equation} 
where $K^{-1}$ is proportional to the average number of collisions per particle $N$, in 
the collision zone of mean transverse size ${\bar R}$; $\lambda$ is the mean free path of 
these particles; $\frac{v_{2}^{\rm{h}}}{\varepsilon_2}$ is the eccentricity-scaled 
flow harmonic expected from perfect fluid hydrodynamics and $K_0$ is a constant estimated to 
be $0.7 \pm 0.03$ with the aid of a transport model \cite{Gombeaud:2007ub}.  
We have found that a modified form of Eq.~\ref{eq2}; 
\begin{equation}
 \frac{v_{2k}}{\varepsilon_{2k}} = \frac{v_{2k}^{\rm{h}}}{\varepsilon_{2k}} \left[ 
  \frac{K^{-1}}{K^{-1}+K_0^{-1}}\right]^{k}\; k=1,2,...\;,	
	\label{eq3}
\end{equation}
provides a good estimate of the scaling violations for all even order harmonics investigated.

Following the operational ansatz of Eq.~\ref{eq3}, an estimate of the extent of 
the deviation from local equilibrium (degree of local equilibrium) can be obtained 
for a given centrality cut as:
\[
\left(\frac{v_{2k}}{v_{2k}^{\rm{h}}}\right)^{1/k} = \frac{1}{(1+K/K_0)}\;\;\; k=1,2,..,
\]
where $K$ is obtained as a function of centrality via a fit to the eccentricity-scaled 
flow coefficients, as discussed below.

\subsection{Coupling strength and the ratio of viscosity to entropy density}\label{viscosity}

	The shear viscosity $\eta$ of the QGP medium reflects its ability to 
flow ``freely'' locally. This medium response to flow gradients is proportional
to the range over which momentum can be readily transported transverse to the flow. 
Consequently, a necessary indication for a strongly coupled QGP would be the observation 
of a rather small value for the ratio of viscosity to entropy 
density $\eta/s$. Here, it is noteworthy that this is a necessary but insufficient 
requirement because a weakly coupled system exhibiting anomalous viscosity could also 
give a relatively small $\eta/s$ value  \cite{Asakawa:2006tc}. The 
ratio $\eta/s$, of course, can not be arbitrarily 
small  \cite{Danielewicz:1984ww,Kovtun:2004de} because 
quantum mechanics limits the size of cross sections via unitarity. 
Therefore, operationally it is the extraction of a relatively small $\eta/s$ value in concert 
with a short mean free path ($\lambda$), from data, that provides a robust indicator 
of a strongly coupled QGP.

For a relativistic fluid, the ratio of viscosity to entropy density can be estimatd as:
\begin{equation}
\frac{\eta}{s}\approx T\lambda c_s \equiv K{\bar R} T c_s,
	\label{eq4}
\end{equation}
where $T$ and $c_s$ are the 
temperature and speed of sound respectively. Thus, the $K$ values extracted from 
fits to eccentricity-scaled flow data, taken in concert with an estimate of $T$ and 
a reliable EOS, can be used to evaluate $\frac{\eta}{s}$.
Similarly, the mean free path ${\lambda}=K{\bar R}$, can be estimated with the aid of 
the centrality dependent geometric value ${\bar R}$.

\section{Proofing the extraction of Knudsen numbers from fits}\label{proofing}
\begin{figure}[!htb]
                 \insertplot{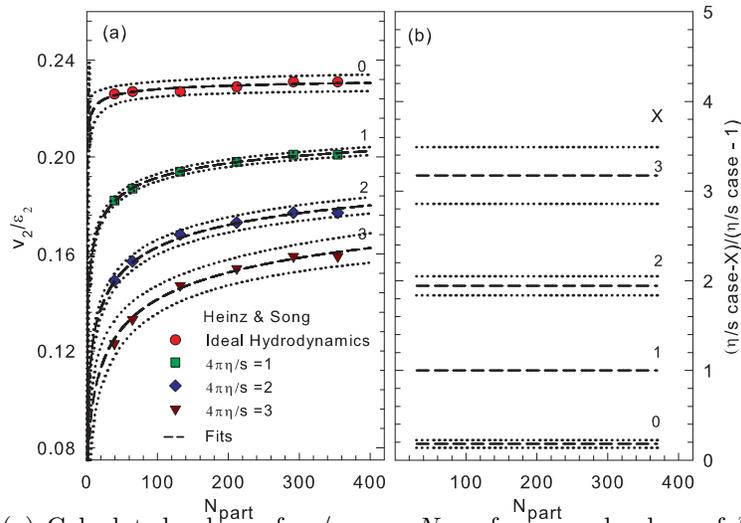}
\vspace*{-1.0cm}
\caption[]{(a) Calculated values of $v_2/\varepsilon_2$ vs. $N_{\rm part}$ for several 
            values of $\frac{\eta}{s}$ \cite{Song:2008si} as indicated. The dashed curves 
            show the fits to these simulated data. (b) $(\frac{\eta}{s}(case-0,1,2,3))/(\frac{\eta}{s}(case-1))$
            vs. $N_{\rm part}$ for for the fits performed in panel (a) see text. The dotted curves 
            represent error bands.
            }
\label{fig1}
\end{figure}
	Reliable estimates for $\frac{\eta}{s}$, $\lambda$ and the degree of local equilibrium 
crucially depend on accurate extractions of the Knudsen number $K$, from fits to eccentricity-scaled 
flow data. 
Therefore, it is important to test the efficacy of the extraction procedure. Such 
a test is illustrated in Fig.~\ref{fig1} where we have performed fits to 
the eccentricity-scaled elliptic flow values $\frac{v_2}{\varepsilon_2}$, recently 
calculated by Song and Heinz \cite{Song:2008si}. 
The symbols in Fig.\ref{fig1}(a) show the $\frac{v_2}{\varepsilon_2}$ values obtained 
from hydrodynamic calculations performed for $4\pi(\frac{\eta}{s})$ values of 0, 1, 2 
and 3 respectively (hereafter referred to as $case-0, 1, 2,$ and $3$ respectively). 
The dashed lines show the corresponding fits obtained with Eq.~\ref{eq3}
and the assumption \cite{Bhalerao:2005mm,Kharzeev:2001gp} that
\[
K^{-1} = \frac{\alpha}{S}\frac{dN}{dy} \sim \beta N_{part}^{1/3}
\]
where $\alpha$ or $\beta$ are fit parameters, $S$ is the  
area of the collision zone, $\frac{dN}{dy}$ is the multiplicity density and 
$N_{part}$ is the number of participants. 

The curves in Fig.\ref{fig1}(a) 
indicate a good fit to the simulated data for each {\it case}. We reiterate 
here that the two parameters of the fit are the scaled hydrodynamic limit 
$\frac{v_{2}^{\rm{h}}}{\varepsilon_2}$ and and $\beta$. The latter allows 
the determination of $K$ for each centrality or value of $N_{\rm part}$.
For each fit indicated in Fig.\ref{fig1}(a), the value of $\frac{v_{2}^{\rm{h}}}{\varepsilon_2}$ 
is within 5\% of the calculated value of 0.23 (cf. solid circles in Fig.\ref{fig1}).	
The dashed lines in Fig.\ref{fig1}(b) show the ratio 
$(\frac{\eta}{s}(case-0,1,2,3))/(\frac{\eta}{s}(case-1))$ determined with Eq.~\ref{eq4} 
and the $K$ values obtained from the extracted values of $\beta$. 
Within errors, the ratios shown in Fig.~\ref{fig1}(b) are the same as the  
ratios of the $\frac{\eta}{s}$ values employed in the hydrodynamic calculations. 
Therefore, we interpret this agreement to be a good validation test of the 
reliability of the the extraction technique.

\begin{figure}[!t]
\vspace*{.4cm}
                 \insertplot{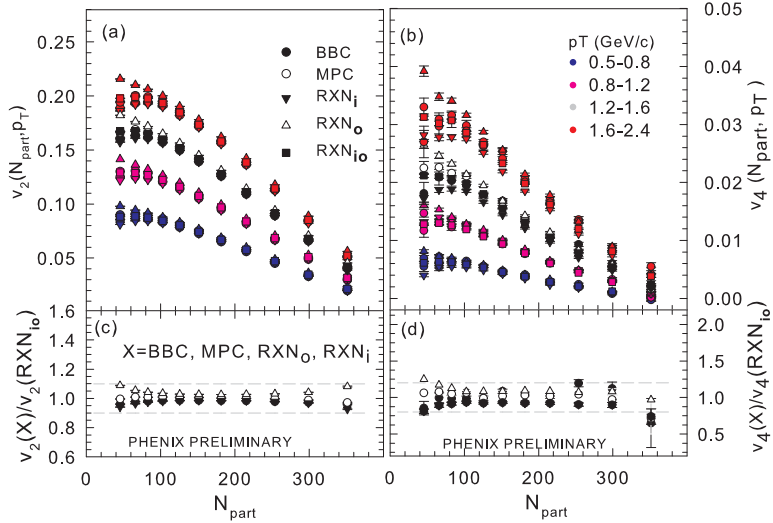}
\vspace*{-0.8cm}
\caption[]{(a) $v_2$ vs. $N_{\rm part}$ and (b) $v_4$ vs. $N_{\rm part}$
for charged hadrons obtained with several reaction plane detectors for the $p_T$ 
selections indicated. The dashed lines in (c) and (d) show 10\% and 20\% error bands
respectively. 
}
\label{fig2}
\end{figure}
\section{Flow measurements and their scaling violations }\label{data}
Reliable estimates for $\frac{\eta}{s}$, $\lambda$ and the degree of local equilibrium also 
demand robust measurements of the flow coefficients. Such measurements of $v_2$ and $v_4$ for charged 
hadrons produced in Au+Au collisions at $\sqrt{s_{NN}}=200$\,GeV, have been recently 
carried out by the PHENIX collaboration. Fig.~\ref{fig2} shows a set of preliminary double 
differential $v_2$  and $v_4$ data obtained from $\sim 3.4 \times 10^9$ minimum-bias 
Au+Au events collected during the 2007 running period. 
These data were obtained via the reaction 
plane method of analysis~\cite{Adler:2003kt};
\begin{equation}
v_{2k} = \frac{\left\langle \cos(2k(\varphi_p -\Phi_2))\right\rangle} 
{\left\langle\cos(2k(\Phi_2 -\Phi_{\rm RP}))\right\rangle}\; k =1,2, 
\label{res}
\end{equation} 
where, $\varphi_p$ is the azimuthal angle  of a charged track
and $\Phi_2$ is the azimuth of the estimated second order reaction (event) plane. 
Five separate event planes were constructed with the aid of the PHENIX Beam-Beam 
Counters (BBC: $3.1 < \left| \eta_{_{\rm BBC}} \right| < 3.9 $), 
Muon Piston Calorimeters (MPC: $3.1 {\,\mathop  < \limits_\sim\,} \left| \eta_{{\rm MPC}} \right| {\,\mathop  < \limits_\sim\,} 3.9 $), 
and the inner (i: $1.5 < \left| \eta_{{\rm RXN_i}} \right| < 2.8$), 
outer (o: $1.0 < \left| \eta_{{\rm RXN_o}} \right| < 1.5$) and combined (io) 
rings  (North and South) of the newly installed PHENIX reaction 
plane detectors (RXN).

The denominator of Eq.~\ref{res} is a resolution 
factor which corrects for the difference between the estimated $\Phi_2$ and 
the true azimuth $\Phi_{\rm RP}$ of the reaction plane \cite{Adler:2003kt}. 
The three sub-event method \cite{Poskanzer:1998yz} was employed to obtain
an estimate of this resolution factor (as a function of centrality) for 
each of the five event planes used in our analysis. For $k=1$, the resolution factor for the 
combined reaction plane from both BBC's has an average of 0.33 over centrality, 
with a maximum $\approx 0.42$ in mid-central collisions~\cite{Adler:2003kt,Adare:2006ti}. 
The MPC and RXN$_{\rm io}$ improve this resolution factor by about 35\% and 
100\%, respectively.  

	For mid-central collisions, a comparison of the double differential flow coefficients 
$v_{2,4}(p_T, N_{\rm part})$, shown for each event plane in Figs.~\ref{fig2}(a) and (b), 	
indicate excellent agreement (i.e much better than 5\% and 10\% for $v_2$ and $v_4$ 
respectively) over the broad range of $p_T$ selections shown. 
For very central and peripheral collisions, this agreement degrades to $\sim$~10\%
and 20\% respectively. Here, it is important to note that;
(i) the $v_4$ signal is rather small (especially for low $p_T$ and central collisions)
and is very difficult to measure precisely with a poor reaction plane resolution. 
(ii) The event plane resolution, for each event plane detector, achieves a maximum 
in mid-central collisions and worsens as one moves towards central 
and peripheral collisions.

	The rather good agreement between the measurements shown in Fig.~\ref{fig2} attests 
to their reliability and to the absence of a significant $\eta$-dependent non-flow 
contribution. A prodigious non-flow contribution, such as from di-jets, would lead to
a sizable difference in the $v_2$ values obtained with event planes determined at 
different rapidity gaps ($\Delta\eta$) with respect to the central arms ~\cite{Jia:2006sb}.

\subsection{Hydrodynamic scaling violations of $v_2$ and $v_4$ }\label{scaling_violations}
	To test for hydrodynamic scaling violations in the data, we divide the $v_2$ and $v_4$ measurements 
obtained with the RXN$_{io}$ event plane\footnote{The RXN$_{io}$ event plane detector provides 
an optimal resolution across collision centralities.} (cf. Fig.~\ref{fig2}), by the respective 
eccentricity (ie. $\varepsilon_2$ and $\varepsilon_4$) for each centrality selection. 
Here, the guiding principle is the prediction from perfect fluid hydrodynamics that 
$v_{2,4}$ is proportional to the initial spatial eccentricity $\varepsilon_{2,4}$ and  
is independent of the size of the collision zone $\bar{R}$. Thus, perfect eccentricity 
scaling would be indicated by a flat dependence for both $\frac{v_2}{\varepsilon_2}$ and 
$\frac{v_4}{\varepsilon_4}$ vs. $N_{\rm part}$.

For each centrality selection, the number of participant nucleons $N_{\rm part}$, 
was estimated  via a Glauber-based model \cite{Miller:2007ri}.
The corresponding transverse size $\bar{R}$ 
and $\varepsilon_{2,4}$ were estimated from the distribution of these 
nucleons in the transverse ($x,y$) plane
via this same Monte-Carlo Glauber model~\cite{Miller:2007ri,Alver:2006wh},  as well
as via the factorized Kharzeev-Levin-Nardi (fKLN) \cite{Drescher:2007ax} model:
\[
\frac{1}{\bar{R}} = \sqrt{\left(\frac{1}{\sigma_x^2}+\frac{1}{\sigma_y^2}\right)},
%
%
\varepsilon_{2} = \frac{\sqrt{(\sigma_y^2-\sigma_x^2)^2 + 4\sigma_{xy}^2}}
{\sigma_x^2+\sigma_y^2},
\varepsilon_{4} = 1 - \frac{8\sigma^2_{xy}}{\sigma^4_x + \sigma^4_y + 2\sigma^2_{xy}},
\]
%
where $\sigma_x$ and $\sigma_y$ are the respective root-mean-square widths of
the density distributions and $\sigma_{xy}=\overline{xy}-\bar{x}\bar{y}$; 
here, bars denote a convolution with the density distribution for a given 
configuration and averaging is performed over configurations. 
This procedure ensures that the fluctuation in the orientation of the initial 
almond-shaped collision zone is taken into account~\cite{Miller:2007ri,Alver:2006wh}.
For the Glauber calculations, the initial entropy profile in the transverse
plane was assumed to be proportional to a linear combination
of the number density of participants and binary collisions \cite{Hirano:2009ah}.
This assures that the entropy density weighting is constrained by multiplicity 
measurements.

\begin{figure}[!t]
\vspace*{.4cm}
                 \insertplot{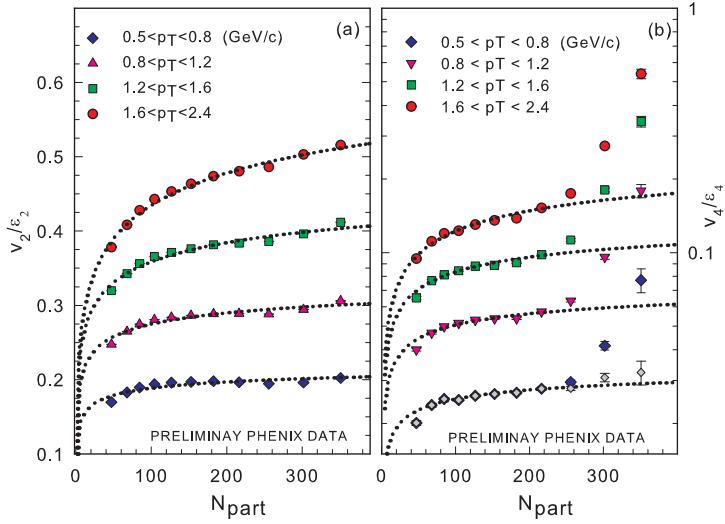}
\vspace*{-0.8cm}
\caption[]{(a) $v_2/\varepsilon_2$ vs. $N_{\rm part}$ and (b) $v_4/\varepsilon_4$ vs. $N_{\rm part}$
for several $p_T$ selections as indicated. The dotted curves are fits to the data with Eq.\ref{eq3}. 
}
\label{fig3}
\end{figure}

		The eccentricity scaled $v_2$ and $v_4$ values obtained with fKLN eccentricities, 
are shown in Fig.~\ref{fig3}. For the lowest $p_T$ selection, they indicate relatively small
scaling violations. However, these violations progressively increase as the $p_T$ for charged hadrons 
is increased. That is, the data points slope progressively upward (from low to high $N_{\rm part}$) 
as the magnitude of the $p_T$ selection is increased.
Relative to mid-central events, a large scaling violation of $\frac{v_4}{\varepsilon_4}$ is 
also apparent for the two most central bins ie. for $N_{\rm part} > 250$.
We have traced this to a small overestimate of fluctuations which only impacts the value 
of $\varepsilon_4$ in the most central collisions as illustrated by the gray diamonds 
in Fig.~\ref{fig3}. The latter shows the resulting values of $\frac{v_4}{\varepsilon_4}$ when 
an attempt is made to account for the overestimate by introducing a very small correlation
between the $\varepsilon_2$ and $\varepsilon_4$ axes.

	A similar scaling test was performed with the Glauber model eccentricities. 
The resulting $\frac{v_2}{\varepsilon_2}$ and $\frac{v_4}{\varepsilon_4}$ values vs. $N_{\rm part}$, 
show trends which are relatively similar to the ones exhibited in Fig.~\ref{fig3}, albeit
with somewhat larger scaling violations. However, as discussed below, our fitting procedure 
provides a good constraint for choosing between the fKLN and Glauber model 
eccentricities. 

\subsection{Estimation of the degree of local equilibrium, $\frac{\eta}{s}$ and $\lambda$ }\label{local_equilibrium}
 
	As discussed earlier, the quantification of scaling violations is  
an important step for reliable estimates of $\frac{\eta}{s}$, $\lambda$ and 
the degree of local equilibrium. This has been accomplished by performing 
simultaneous fits to the eccentricity-scaled $v_2$ and $v_4$ data for each 
of the $p_T$ selections. For these fits, we follow the same fitting ansatz proofed 
in section \ref{proofing}, ie. we use Eq.~\ref{eq3};
\[
 \frac{v_{2k}}{\varepsilon_{2k}} = \frac{v_{2k}^{\rm{h}}}{\varepsilon_{2k}} \left[ 
  \frac{K^{-1}}{K^{-1}+K_0^{-1}}\right]^{k}\; k=1,2 \;\; {\rm and }\;\; 
   K^{-1} = \beta N_{part}^{1/3}.
\]
Here again, the scaled hydrodynamic limits $\frac{v_{2k}^{\rm{h}}}{\varepsilon_{2k}}$, and $\beta$ are fit 
parameters. Note as well that $\beta$ allows the extraction of $K$ values as a 
function of $N_{\rm part}$, for each fit.

	The requirement of a simultaneous fit to the eccentricity-scaled data  
ensures that the same $K$ values are extracted from the $v_2$ and $v_4$ measurements.
Equally important is the fact that they provide a constraint for making a choice between 
the fKLN and Glauber eccentricities. That is, Glauber eccentricities result in relatively  
poor simultaneous fits while the fKLN eccentricities give very good simultaneous fits 
($R_{\rm adj}^2 \sim 1$) as shown by the dotted curves in Fig.~\ref{fig3}. These fits 
underscores the fact that data comparisons for different centralities, as well as 
ratios such as $\frac{v_4}{(v_2)^2}$ must take account of eccentricity differences.
 
As pointed out earlier, each fit gives a value for the hydrodynamic limits 
$\frac{v^h_{2}}{\varepsilon_{2}}$ and $\frac{v^h_{4}}{\varepsilon_{4}}$,
and a $\beta$ value which determines $K$ as a function of $N_{\rm part}$. 
With these $K$ values, the degree of local equilibrium and $\frac{\eta}{s}$ can be estimated 
using Eqs.~\ref{eq3} and \ref{eq4} respectively. To estimate $\frac{\eta}{s}$
at a given centrality, we assume a temperature  
$T = 220\pm 20$~MeV \cite{Adare:2008fqa} for the plasma when flow develops\footnote{A centrality 
dependent $T$ would results in an additional uncertainty on the 
extracted $\frac{\eta}{s}$ values.}, 
and use the the lattice EOS to estimate 
the associated value of $c_s = 0.47 \pm 0.03$~c. For any given 
centrality, $\lambda = \bar{R}K$ is evaluated with the aid of the 
calculated transverse size $\bar{R}$, obtained for that centrality.

\begin{figure}[!t]
\vspace*{.4cm}
                 \insertplot{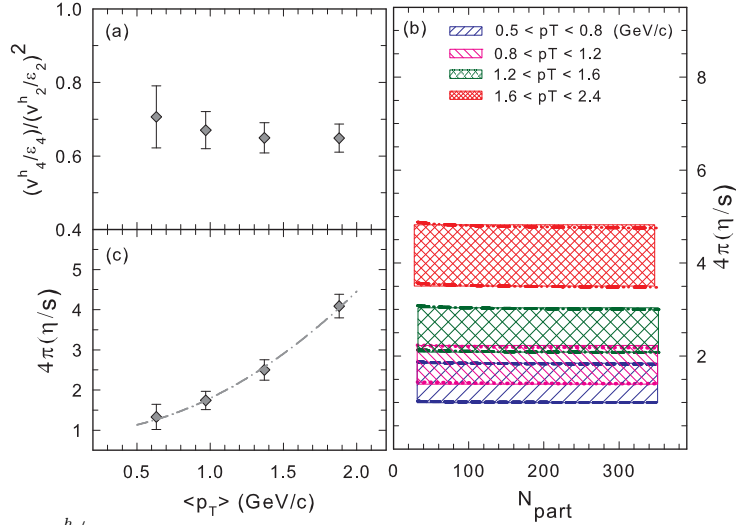}
\vspace*{-0.8cm}
\caption[]{(a) $\frac{v^h_4/\varepsilon_4}{(v^h_2/\varepsilon_2)^2}$ vs. $p_T$, see text; (b) extracted 
bands for $4\pi(\eta/s)$ vs. $N_{\rm part}$ for several $p_T$ selections as 
indicated; (c) extracted values of $4\pi(\eta/s)$ vs. $p_T$. 
}
\label{fig4}
\end{figure}

	Figure~\ref{fig4} summarize the main results from our extractions with the 
fKLN eccentricities. It is noteworthy that the trends of the results 
obtained with the Glauber model eccentricities are essentially the same, 
albeit with different magnitudes. 
Panel (a) shows that, within errors, the ratio of the extracted hydrodynamic 
limits $\frac{(v^h_4/\varepsilon_4)}{(v^h_2/\varepsilon_2)^2}$ change little, if any, 
with hadron $\left<p_T\right>$. The extracted bands for $4\pi(\eta/s)$ are shown 
as a function of $N_{\rm part}$ in Fig.~\ref{fig4}(b). The essentially flat dependence on 
$N_{\rm part}$ demonstrates the important role of the transverse size and 
the eccentricity of the collision zone. It further suggests that 
even though $K$ increases as collisions become more peripheral, $\lambda$ 
does not have a strong dependence on collision centrality. 

	In contrast to the $N_{\rm part}$ dependence, Fig.~\ref{fig4}(b) implies
a relatively strong $p_T$ dependence for the extracted $4\pi(\eta/s)$ values, ie.
$\eta/s$ increases by almost a factor of four over the indicated $p_T$ range.
This $p_T$ dependence is made more transparent in Fig.~\ref{fig4}(c) where 
we plot $4\pi(\eta/s)$ vs. the mean value of the $p_T$ ranges indicated 
in Fig.~\ref{fig4}(b). The dashed-dot curve in the figure illustrates 
the quadratic nature of this $p_T$ dependence. Here, we wish to emphasize that this 
quadratic dependence on $p_T$ is {\bf not} related to an intrinsic property of the QGP. Instead, it 
reflects the finding by Teaney \cite{Teaney:2003kp} that viscous corrections to ideal 
hydrodynamics grow as
\[
   \left(\frac{p_T}{T}\right)^2 K.
\]
Since our extracted $K$ values include the $p_T$-dependent factor given in the 
above expression, they give rise to the quadratic $p_T$ dependence of $4\pi(\eta/s)$ observed  
in Figs.~\ref{fig4}(b) and (c). This also indicate that a $p_T$-independent estimate 
of $4\pi(\eta/s)$ is obtained when the magnitude of $p_T \approx T$. 

	The slow rate of increase of $4\pi(\eta/s)$ at low $p_T$ (cf. Fig.~\ref{fig4}(c)) 
allows the use of the results for lowest $p_T$ hadrons to constrain the 
estimates $4\pi(\eta/s) = 1.3 \pm 0.3$ and 
$\lambda = 0.25 - 0.3$~fm for the plasma. The corresponding estimate for the 
degree of local equilibrium is within 5~-~10\% of the value for a system having 
an $\frac{\eta}{s}$ value equal to the conjectured lower bound of $\frac{1}{4\pi}$.
To estimate the latter, we use the $K$ values extracted from the fit to 
the simulated data (for $4\pi(\eta/s)=1$) shown in Fig.~\ref{fig1}.
Our $\eta/s$ estimate is in good agreement with prior 
extractions~\cite{Shuryak:2008eq,Romatschke:2007mq,Xu:2007jv,Drescher:2007cd,Lacey:2006bc,Adare:2006nq}.
Our $\lambda$ estimate also indicate that the low $\eta/s$ value  
is associated with a relatively short mean free path in the plasma. 
We interpret these observations as an important indication for a strongly coupled QGP.

\section{Conclusions}\label{concl}

To conclude, we have made detailed studies of possible hydrodynamic scaling violations
of the eccentricity scaled $v_4$ and $v_2$ flow coefficients for charged hadrons.
These studies validate the hydrodynamic scaling patterns 
expected for a nearly inviscid system close to thermal equilibrium 
ie. only 5-10\% less than the value estimated for a system having an $\eta/s$ 
value equal to the conjectured lower bound of $1/4\pi$. They also provide
the estimates $4\pi(\eta/s) = 1.3 \pm 0.3$ and 
$\lambda = 0.25 - 0.3$~fm. These estimates suggest that the transverse 
expansion dynamics leading to anisotropic flow favor the creation of a 
strongly coupled QGP in collision zones for central and 
mid-central Au+Au collisions at $\sqrt{s_{NN}}~=~200$ GeV.
 
\section*{Acknowledgments}
This work was supported by the US DOE under
contract DE-FG02-87ER40331.A008.
 
\begin{notes}
\item[a]
Permanent address: Chemistry Dept., Stony Brook University, Stony Brook, USA;\\ 
E-mail: Roy.Lacey@Stonybrook.edu
\end{notes}
 
\bibliography{bigsky2009_Refs_Lacey}
\bibliographystyle{bigsky2009}
 
\vfill\eject
\end{document}